\documentclass[preprintnumbers,nofootinbib,superscriptaddress,10pt]{revtex4}
\usepackage[dvipdfmx]{graphicx} 
\usepackage{amsmath,amssymb,amsfonts,bm} 

\def\a{\alpha}  \def\g{\gamma}  \def\d{\delta}     \def\th{\theta}     \def\m{\mu} \def\n{\nu}     \def\r{\rho}   \def\t{\tau}       

\def\dg{\dagger}  \def\nn{\nonumber}

\newcommand{\lsp}{ \left ( } \newcommand{\rsp}{ \right ) }   \newcommand{\To}{\Rightarrow}

\def\abs#1{\left| #1\right|}

\newcommand{\Diag}[3]{ \begin{pmatrix} #1 & 0 & 0 \\ 0 & #2 & 0 \\ 0 & 0 & #3 \\\end{pmatrix}}

\usepackage{color}

\begin{document}

\title{Explicit Rephasing Transformation to PDG Parameterization and \\ 
Simplified Expression of the Dirac CP Phase by Fermion-Specific Invariants}

\preprint{STUPP-25-289}

\author{Masaki J. S. Yang}
\email{mjsyang@mail.saitama-u.ac.jp}
\affiliation{Department of Physics, Saitama University, 
Shimo-okubo, Sakura-ku, Saitama, 338-8570, Japan}
\affiliation{Department of Physics, Graduate School of Engineering Science,
Yokohama National University, Tokiwadai, Hodogaya-ku, Yokohama, 240-8501, Japan}



\begin{abstract} 

In this letter, we present an explicit rephasing transformation
that maps an arbitrary mixing matrix $U$ to the PDG standard parameterization
 $ U^{0} =  {\rm diag} ( e^{ - i \arg U_{e1} }  \, , \, e^{ i  \arg [ {U_{e2}  U_{\tau 3} \over \det U  } ]} \, , \, e^{ i \arg [ {  U_{e2} U_{\mu 3} \over \det U } ] } )$ $U \, {\rm diag} ( 1 \, , \, e^{ - i \arg [{U_{e2} \over U_{e1} } ] }  \, , \, e^{ - i \arg [ { U_{e2} U_{\mu 3} U_{\tau 3} \over \det U } ] })$. 
By this procedure, which has remained largely conceptual for more than four decades, 
 six independent phases of the mixing matrix are expressed systematically in terms of the arguments. 
We also apply this framework to the fermion diagonalization matrices $U^{\nu,e}$ under the approximation $U^{\nu,e}_{13}=0$. 
By partially employing the inversion formula, we factorize all CP phases through the explicit rephasing, thereby revealing the entire CP structure of the mixing matrix.
The observable Dirac phase depends only on two relative phases between the fermion sectors and is determined by fermion-specific invariants.

\end{abstract} 

\maketitle

\section{Introduction}

The CP-violating phase appearing in the mixing matrix directly encodes an essence of flavor-dependent CP violation in  the Standard Model. 
Observations of the Dirac CP phase in the lepton mixing matrix are ongoing \cite{NOvA:2021nfi, T2K:2021xwb}, and more precise measurements are planned for the future \cite{DUNE:2020jqi, Hyper-KamiokandeProto-:2015xww}.
Indirect probes of the Majorana phases are also continuing through experiments on the neutrinoless double beta decay \cite{KamLAND-Zen:2022tow}. 
Since the advent of the Jarlskog invariant \cite{Jarlskog:1985ht}, 
the magnitude of CP violation has been extensively studied using imaginary parts of rephasing invariants
\cite{Wu:1985ea, Bernabeu:1986fc, Gronau:1986xb, Branco:1987mj, Bjorken:1987tr, Nieves:1987pp, Botella:1994cs,  Kuo:2005pf, Jenkins:2007ip, Branco:2008ai, Chiu:2015ega}. 

To extract the CP phase of the mixing matrix, it is customary to state that one can transform the mixing matrix to the PDG parameterization by a rephasing transformation at a conceptual level \cite{Xing:2002sw, Antusch:2005kw, Farzan:2006vj, Hochmuth:2007wq, Ge:2011qn, Marzocca:2013cr,  Petcov:2014laa, Dasgupta:2014ula}.
Despite more than forty years since the original work of Chau and Keung~\cite{Chau:1984fp},
almost no study has explicitly presented such a transformation.

In contrast, recent studies show that the Dirac CP phase is expressed directly
in terms of rephasing invariants involving the determinant of mixing matrix 
\cite{Yang:2025hex,Yang:2025cya,Yang:2025law,Yang:2025ftl,Yang:2025dhm,Yang:2025vrs}. 
Although the determinant has been historically omitted as an unphysical phase, 
 it implicitly retains global phase information relevant for reconstructing CP phases. 
However, these previous studies of the Dirac phase $\delta$, 
 together with earlier conventional approaches based on the $\sin \d$-type invariants, 
 did not directly elucidate the entire phase structure of the mixing matrix. 

Therefore, to fill this long-standing gap, 
we present an explicit rephasing transformation that maps a general unitary matrix to the PDG standard form.
Six independent phases, including the Dirac and Majorana phases, are systematically derived from matrix elements and the determinant. 
We further extend this framework to the diagonalization matrices of fermions $U^{\n , e}$,
allowing the observable CP phases to be expressed by fermion-specific rephasing invariants of two fermion sectors.

\section{Explicit rephasing transformation to PDG standard form}

Here, we perform the explicit rephasing transformation by deriving the Dirac phase $\delta$ and the Majorana phases $\alpha_{2,3}$.
The standard PDG parameterization of the lepton mixing matrix is given by \cite{Chau:1984fp, ParticleDataGroup:2018ovx} 
\begin{align}
U^{\rm PDG} &= U^{0} P \equiv
\begin{pmatrix}
c_{12} c_{13} & s_{12} c_{13} & s_{13} e^{-i\d} \\
-s_{12} c_{23} - c_{12} s_{23} s_{13}  e^{i \d} & c_{12} c_{23} - s_{12} s_{23} s_{13} e^{i \d} & s_{23} c_{13} \\
s_{12} s_{23} - c_{12} c_{23} s_{13} e^{i \d} & -c_{12} s_{23} - s_{12} c_{23} s_{13} e^{ i \d} & c_{23} c_{13}
\end{pmatrix}  
\Diag{1}{e^{ i \a_{2} / 2}}{e^{ i \a_{3} / 2}} .
\label{PDG}
\end{align}
By using rephasings of five fermion fields, together with the sign freedom of $\delta$,
we  choose all mixing angles to be positive, $c_{ij}, s_{ij} > 0$.
For a lepton mixing matrix  $U$ defined in an arbitrary basis, 
the rephasing transformation between $U$ and $U^{0}$ is written in the following form: 
\begin{align}
\begin{pmatrix}
U_{e1} & U_{e2} & U_{e3}  \\
U_{\m1} & U_{\m 2} & U_{\m 3}  \\
U_{\t 1} & U_{\t 2} & U_{\t 3}  \\
\end{pmatrix}
= 
\Diag{e^{i \g_{L1}}}{e^{i \g_{L2}}}{e^{i \g_{L3}}}
\begin{pmatrix}
|U_{e1}| & |U_{e2}| & |U_{e3}| e^{-i\d} \\
U_{\m1}^{0} & U_{\m 2}^{0} & |U_{\m 3}|  \\
U_{\t 1}^{0} & U_{\t 2}^{0} & |U_{\t 3}|  \\
\end{pmatrix}
\Diag{e^{ - i \g_{R1}}}{e^{ - i \g_{R2}}}{e^{ - i \g_{R3}}} . 
\end{align}

The Dirac phase is determined by solving the unknown phases $\gamma_{Li}$ and $\gamma_{Ri}$
 from the determinant and matrix elements whose arguments are trivial:
\begin{align}
|U_{e3}| e^{i \d } = U_{e3}^{*}e^{i(\g_{L1} - \g_{R3} )}  \, , ~~ 
 \arg \left[ {U_{e1} U_{e2} U_{\m3} U_{\t 3} \over \det U } \right] & = 
 \g_{L1}  -  \g_{R3} 
 ~~ \To ~~ 
\d = \arg \left[ {U_{e1} U_{e2} U_{\m3} U_{\t 3}   \over U_{e3} \det U } \right] . 
\label{Dirac}
\end{align}
Including the transformation property of the determinant, 
this expression is manifestly rephasing-invariant and thus corresponds to a physical observable. 
Its equivalence to the Jarlskog invariant $J$ can be readily shown \cite{Yang:2025hex}: 
\begin{align}
 \sin \arg \left[ {U_{e1} U_{e2} U_{\m3} U_{\t 3}   \over U_{e3} \det U } \right] =  \frac{ 1 - |U_{e 3}^{2}|  }{ |U_{e1} U_{e2} U_{\m 3} U_{\t 3}  U_{e3}| } \, J = \sin \d \, . 
\end{align}
An advantage of expressing CP phases in terms of arguments is the simplicity of formulation. 
In the conventional description based on imaginary parts, a sum of phases inevitably lead to lengthy products of trigonometric functions, which obscure the dependence on underlying phases.

The phase differences $\gamma_{R1}-\gamma_{R2,3}$ reproduce the well-known results of the Majorana phases  \cite{Doi:1980yb}, 
\begin{align}
 \arg [ U_{e1}^{*} U_{e2}]& = \g_{R1} - \g_{R2} = \a_{2} / 2 \, , ~~~
\arg [ U_{e1}^{*} U_{e3}]  = \g_{R1} - \g_{R3} - \d = \a_{3} / 2 - \d\, ,
\label{Majorana}
\end{align}
because the phases $\a_{2,3}$ are defined in the basis where $\g_{R1} = 0$. 
By combining the phase with $\delta$, we directly obtain an explicit expression for $\alpha_3/2$; 
\begin{align}
{\a_{3} \over 2} = \g_{R1} - \g_{R3} = \arg \left[ {U_{e3} \over U_{e1} } \right]  +  \arg \left[ {U_{e1} U_{e2} U_{\m3} U_{\t 3}   \over U_{e3} \det U } \right] 
=  \arg \left[ {U_{e2} U_{\m3} U_{\t 3}   \over  \det U } \right] . 
\end{align}

This observation naturally motivates us to express the unphysical phases $\gamma_{Li}$ 
 in terms of the arguments of matrix elements.
As the remaining three conditions, we choose
\begin{align}
\g_{L1} - \g_{R1} = \arg U_{e1} \, , ~~
\g_{L2} - \g_{R3} = \arg U_{\m3} \, , ~~
\g_{L3} - \g_{R3} = \arg U_{\t 3} \, . 
\end{align}
By the freedom of an overall phase,
one of $\gamma_{Li}$ and $\gamma_{Ri}$ remains undetermined. 
Leaving $\gamma_{R1}$ as a free parameter to preserve the form of the Majorana phases, 
the phase $\gamma_{R1}$ cancels between the left and right phase matrices. 
%
%
As a result, the rephasing from $U$ to $U^{0}$ is obtained as 
\begin{align}
U^{0} & = 
\Diag {e^{ - i \arg U_{e1} } }{e^{ i  \arg \left[ {U_{e2}  U_{\t 3} \over \det U  } \right]}}{e^{ i \arg \left[ {  U_{e2} U_{\m3} \over \det U } \right] }}  U 
\Diag{1}  {e^{ - i \arg \left[{U_{e2}\over U_{e1} } \right] } } {e^{ - i \arg \left[ { U_{e2} U_{\m3} U_{\t 3} \over \det U } \right] }} .  \label{exptrf}
\end{align}
It provides an explicit realization of the rephasing transformation to the PDG parameterization,
which has remained largely conceptual for nearly half a century.
Alternatively, without explicitly substituting $\delta$,  one ultimately obtains,
\begin{align}
U^{0} & = 
\Diag {1 }{e^{  i \d + i \arg [U_{e3} / U_{\m 3} ]}  }{e^{  i \d + i \arg [U_{e3} / U_{\t 3} ]} }  U 
\Diag{e^{ - i \arg U_{e1} } }{e^{ - i \arg U_{e2}  }} {e^{ - i \d - i \arg  U_{e3} }} . 
\end{align}
In this way, the procedure renders the arguments of the first row and the third column trivial, while setting the argument of the (1,3) element to $-\delta$.

Through this explicit rephasing, all six independent phases of a unitary matrix are expressed in terms of the arguments of its matrix elements and the determinant.  
The remaining four nontrivial arguments are expressed by other rephasing invariants 
\begin{align}
\arg U_{\m1}^{0} & = \arg \left[ { U_{e2} U_{\m1} U_{\t 3} / \det U } \right] \, , ~~ 
\arg U_{\m 2}^{0} = \arg \left[ { U_{e1} U_{\m2} U_{\t 3} / \det U } \right] \, , \nn \\
\arg U_{\t 1}^{0} & = \arg \left[ { U_{e2} U_{\m3} U_{\t1} /  \det U } \right] \, , ~~ 
\arg U_{\t 2}^{0}  = \arg \left[ { U_{e1} U_{\m3} U_{\t2} /  \det U } \right] \, .  \label{third}
\end{align}
The validity of these relationships is manifest because $U_{e1}, U_{e2}, U_{\m3}, U_{\t3}$ and $\det U$ are real in the PDG parametrization. 

By applying the explicit rephasing transformation to the diagonalization matrices  of neutrinos $U^{\nu}$ and 
charged leptons $U^{e}$, 
we identify more fundamental CP phases underlying the lepton mixing matrices. 
Regarding the diagonalization matrices $U^{\n , e}$, 
their explicit rephasings are defined as 
$ U^{\nu,e} = \Phi^{\nu,e}_{L}\, U^{\nu,e 0}\, \Phi^{\nu,e}_{R}, $
where $U^{\nu,e 0}$ are their PDG standard form and
$\Phi^{\nu,e}_{L,R}$ are phase matrices expressed in terms of matrix elements.
It then follows that the mixing matrix 
$U \equiv U^{e\dagger} U^{\nu}$ is 
\begin{align}
U = \Phi^{e \dagger}_{R} \, U^{e 0\dagger} \, \Phi_L^{e\dagger} \Phi_L^\nu \, U^{\nu 0} \, \Phi_R^{\n} \, .
\end{align}
As sources of physical CP violation,
the Dirac-like phases $\delta^{\nu, e}$ contained in $U^{\nu,e0}$
are given by the rephasing invariants 
\begin{align}
\d^{\n , e} =  \arg \left[ {{U_{11}^{\n , e} U_{12}^{\n , e} U_{23}^{\n , e} U_{3 3}^{\n , e} \over U_{13}^{\n , e} \det U^{\n , e} } }\right] . 
\end{align}

Among the remaining CP phases, we combine the left-handed phase matrices and define
$\Phi_{L} \equiv \Phi_{L}^{e \dagger}\Phi_{L}^{\n} \equiv \mathrm{diag}\,( e^{i\rho_{1}},\, e^{i\rho_{2}},\, e^{i\rho_{3}} )$.
The relative phases between the neutrinos and charged leptons are found to be
\begin{align}
\r_{1} = \arg \left[ { U^{e *}_{11} U^{\n}_{11} } \right] , ~~
\r_{2} =  \arg \left[ { \det U^{e *} \over U^{e *}_{12}  U^{e *}_{33} } { \det U^{\n} \over U^{\n}_{12}  U^{\n}_{33} } \right] , ~~ 
\r_{3} = \arg \left[ {  \det U^{e *} \over U^{e *}_{12} U^{e *}_{23} } { \det U^{\n} \over U^{\n}_{12} U^{\n}_{23} } \right] . 
\end{align}
These phases represent the relative misalignment of underlying CP phases between fermion sectors.
Because of the freedom of an overall phase, 
only the relative phases $\rho_i - \rho_{j}$ between $\r_{i}$ are physically meaningful. 
Two of $\rho_i - \rho_{j}$ are independent, and they are invariant under a rephasing transformation of $U = U^{e \dagger} U^{\nu}$. 

When any matrix element $U^{f}_{ij}$  vanishes, the corresponding phase $\d^{f}$ becomes zero, and the phase structure of $U^{f}$  is simplified.
In this case, the phases of all matrix elements are removed by the rephasing transformations, 
and vanishing element $U^{f}_{ij}$ in $\r_{i}$ can be replaced by other matrix elements with the same rephasing property. 
For example, in the case of $U_{12}^{e} = 0$, partially applying the inversion formula except in the first row and the second column, one obtains 
\begin{align}
U' = 
\begin{pmatrix}
U^{e}_{11}  & 0 & U^{e}_{13} \\[2pt]
{ U^{e*}_{13} U^{e*}_{32} \over \det U^{e *}} & U^{e}_{22} & - {U_{11}^{e *} U_{32}^{e *} \over \det U^{e *}} \\[2pt]
- {U^{e *}_{13} U^{e *}_{22} \over \det U^{e *}} & U^{e}_{32}& {U_{11}^{e *} U_{22}^{e *} \over \det U^{ e *}} \\
\end{pmatrix} . 
\end{align}
All CP phases in this matrix can be removed by a rephasing transformation 
\begin{align}
& {\rm diag} ({e^{ - i \arg U^{e}_{11} } } \, , \, {e^{ i  \arg \left[ {U_{13}^{e}  U_{32}^{e} \over \det U^{e}  } \right]}}  \, , \, {e^{ i \arg \left[ {  U_{13}^{e} U_{22}^{e} \over \det U^{e} } \right]}}  ) \, U'  \,
{\rm diag} ({1} \, , \, {e^{ - i \arg \left[{U^{e}_{13} U^{e}_{22} U^{e}_{32}\over \det U^{e} } \right] } } \, , \, {e^{ - i \arg \left[ { U^{e}_{13} \over U_{11}^{e}} \right] }}) 
= |U'| \, , 
\end{align}
where $|U'|$ represents a matrix in which all elements are real.
As a result, finite relative phases $\rho_{2,3}$ are obtained through the following replacement.
\begin{align}
\arg \left[ { U^{e}_{12}  U^{e}_{33} } \right] \to  \arg \left[ { U^{e}_{13}  U^{e}_{32} } \right] 
, ~~ 
\arg \left[ {  U^{e}_{12} U^{e}_{23} } \right] \to \arg \left[ {  U^{e}_{13} U^{e}_{22} } \right] .
\end{align}
This replacement is not allowed when $\d^{e} \neq 0$, because the nontrivial phases in the PDG parametrization in Eq.~(\ref{third}) cannot be eliminated. 

The Majorana phases are also functions of these underlying phases. 
Since $U^{f0}$ is written in terms of the Dirac-like phase $\d^{f}$ and mixing angles, 
the Majorana phases $\alpha_{2,3}$ are expressed as functions of five phases: 
the two Dirac-like phases $\delta^\nu$ and $\delta^e$, two relative phases $\r_{j} - \r_{k}$, 
and each Majorana-like phase $\arg [U^\nu_{1i}/U^\nu_{11} ]$.

If the mixings of $U^{e}$ are as small as those of the CKM matrix owing to some grand unification relationship,  
the term involving $U^{e}_{31}$ is neglected and the Majorana phases are approximated as 
\begin{align}
{\a_{2} \over 2} & \simeq \arg \left [{ U^{\n}_{12} \over U^{\n}_{11}} \right ]
 + \arg \left[ \frac{ |U^{e}_{11} U^{\n}_{12}| + e^{ i (\r_{2} - \r_{1})} U^{e 0 *}_{21} U^{\n 0}_{22} }
 {|U^{e}_{11} U^{\n}_{11}| + e^{ i (\r_{2} - \r_{1}) } U^{e 0 *}_{21} U^{\n 0}_{21} } \right] , \nn \\
{\a_{3} \over 2} - \d & \simeq \arg \left[ {U_{13}^{\n} \over U_{11}^{\n } } \right ] + 
\arg  \left[ { \abs {U^{e}_{11} U^{\n}_{13}} +  e^{i (\d^{\n} + \r_{2} - \r_{1})}  U^{e 0 *}_{21} |U^{\n}_{23}|  \over 
\abs {U^{e}_{11} U^{\n}_{11}} + e^{i (\r_{2} - \r_{1}) } U^{e 0 *}_{21} U^{\n 0}_{21}  }  \right] . 
\end{align}
In this case, the phases $\a_{2}/2$ and $\a_{3}/2 - \d$ are reduced to functions of four underlying phases. 
This decomposition clarifies which CP phases originate from each fermion sector
and  from their relative misalignment. 
A similar representation can be obtained for the Dirac phase, which is a function of four fundamental phases, 
$\delta^\nu$, $\delta^e$, $\rho_2-\rho_3$, and $\rho_3-\rho_1$.  

\section{Simplified expression of the Dirac CP phase by fermion-specific invariants}

In this section, we investigate the behavior of the Dirac phase in a simplified situation.
For the lepton mixing matrix $U \equiv U^{e \dg} U^{\n}$ with the diagonalization matrices $U^{\n , e}$, 
we introduce the following approximation.

\begin{description}
\item[\bf Approximation:] 
The (1,3) elements $U_{13}^{\n, e}$ of $U^{\n,e}$ are treated as negligible. 
More general cases will be perturbatively calculated from this limit.

\item[\bf Justification:] 

When the mass matrices $m_{\n, e}$ possess chiral symmetries for the first and second generations,
$m_{\n, e} = D_{L} \, m_{\n, e} \, D_{R}$, all mixings and lighter singular values vanish. 
Here, $D_{L,R}$ are defined as $ D_{L,R} \equiv {\rm diag} \, (e^{i\phi_{L,R}^{1}}, \, e^{i\phi_{L,R}^{2}}, \, 1 )$  
with phases $\phi_{L,R}^{1,2}$. In realistic situations, these chiral symmetries are only approximate, 
and the mixing angles are suppressed by powers of ratios of the singular values $m_{fi}/m_{fj}$. 
In the actual mixing matrix, $|U_{e3}| \simeq 0.15$ is not particularly small. 
However, as shown in the next subsection, if the magnitude of $|U^{e}_{12}|$ is close to the Cabibbo angle $\sin \th_{C} \sim 0.2$, 
the realistic value $|U_{e3}| \simeq |U^{e}_{12} U^{\nu}_{23}| \simeq 0.15$ is obtained from the bi-maximal mixing $|U^{\nu}_{23}| \sim 0.7$. 

\end{description}

Under this approximation, let us express the Dirac phase in terms of fermion-specific rephasing invariants.
To implement the unitarity constraints, the matrix inversions are partially applied as 
\begin{align}
U &= 
\begin{pmatrix}
U_{11}^{e *} & - {U_{12}^e U_{33}^e \over \det U^{e}} & { U_{12}^{e} U_{23}^{e} \over \det U^{e}}  \\[2pt]
U_{12}^{e *} & {U_{11}^e U_{33}^e \over \det U^{e} } & - {U_{11}^{e} U_{23}^{e} \over \det U^{e } } \\[2pt]
0 & U_{23}^{e *} & U_{33}^{e *} \\
\end{pmatrix} 
\begin{pmatrix}
U_{11}^{\n} & U_{12}^{\n} & 0\\[2pt]
- {U_{12}^{\n*} U_{33}^{\n*} \over \det U^{\n *} }  & {U_{11}^{\n*} U_{33}^{\n*} \over \det U^{\n*}} & U_{23}^{\n} \\[2pt]
{ U_{12}^{\n*} U_{23}^{\n*} \over \det U^{\n *}} & - {U_{11}^{\n*} U_{23}^{\n*} \over \det U^{\n * } } & U_{33}^{\n} \\
\end{pmatrix} \nn  \\ 
& = 
\begin{pmatrix}
\tilde U_{e1} & \tilde  U_{e2}  &  - {U_{12}^e \over \det U^{e} }  (U^{e}_{33} U^{\n}_{23} - U^{e}_{23} U^{\n}_{33} )  \\[2pt]
* & * & {U_{11}^e \over \det U^{e} }  (U^{e}_{33} U^{\n}_{23} - U^{e}_{23} U^{\n}_{33} )  \\[2pt]
* & * & U^{e *}_{23} U^{\n}_{23} +  U^{e *}_{33} U^{\n}_{33}
\end{pmatrix} . \label{U}
\end{align}
Here, the matrix elements denoted by $*$ are not important for the phase calculation, and 
\begin{align}
\tilde U_{e1}  & = U_{11}^{e *} U_{11}^{\n} + {U_{12}^e U_{33}^e \over \det U^{e}}{U_{12}^{\n*} U_{33}^{\n*} \over \det U^{\n *} } + { U_{12}^{e} U_{23}^{e} \over \det U^{e}} { U_{12}^{\n*} U_{23}^{\n*} \over \det U^{\n *}}  \, ,  \nn \\
\tilde U_{e2} & = U_{11}^{e *} U_{12}^{\n}  - {U_{12}^e U_{33}^e \over \det U^{e}} {U_{11}^{\n*} U_{33}^{\n*} \over \det U^{\n*}} - { U_{12}^{e} U_{23}^{e} \over \det U^{e}} {U_{11}^{\n*} U_{23}^{\n*} \over \det U^{\n * } } \, .
\end{align}
Since the factor $ (U^{e}_{33} U^{\n}_{23} - U^{e}_{23} U^{\n}_{33} ) / \det U^{e}$ in Eq.~(\ref{U}) cancel out, 
the CP phase is found to be
\begin{align}
\d & = 
\arg \bigg[ {\tilde U_{e1} \over \tilde U_{e2}^{*}} \bigg]
 + \arg \bigg[ - \frac{ U^{e}_{11} ( U^{e *}_{23}U^{\n}_{23} + U^{e *}_{33}U^{\n}_{33}  ) }
{U^{e}_{12} \det U^{\n } \det U^{e *} } \bigg ] . 
\end{align}

On the other hand, the approximation leads to $\d^{\n,e} = 0$, and 
all CP phases are removed by the explicit rephasing~(\ref{exptrf}).  
Thus, only two relative phases between $U^{\n}$ and $U^{e}$ contribute to the physical CP phases.
To demonstrate this point, we define the following two ratios 
\begin{align}
R_{12} \equiv  { U^{e *}_{11}  U^{e *}_{12}  U^{e *}_{33} \over \det U^{e *}  } { U^{\n}_{11} U^{\n}_{12}  U^{\n}_{33} \over \det U^{\n} } \, ,  ~~
R_{23} \equiv { U^{e *}_{23} U^{\n}_{23}   \over U^{e*}_{33} U^{\n}_{33}  } \, , ~~ 
R_{12} R_{23} = 
 { U^{e *}_{11} U^{e *}_{12} U^{e *}_{23} \over \det U^{e *} } { U^{\n}_{11} U^{\n}_{12} U^{\n}_{23} \over \det U^{\n} } \, . 
\end{align}
Then the Dirac phase is written as a function of $R_{12}$ and $R_{23}$ 
\begin{align}
\d
& = \arg \left [ 
 \dfrac{ |U_{11}^{e} U_{11}^{\n}|^{2}  + R_{12}^{*} + R_{12}^{*}  R_{23}^{*} }
{ \lsp |U_{11}^{e} U_{12}^{\n}|^{2} - R_{12}^{*} - R_{12}^{*} R_{23}^{*} \rsp^{*} } \right]
 + \arg \bigg[ - { U^{e *}_{11} U^{\n}_{12}  \over  U^{e}_{12}  U^{\n *}_{11} }  \frac{ U^{e *}_{33}U^{\n}_{33}  (  1 + {U^{e *}_{23}U^{\n}_{23} \over U^{e *}_{33}U^{\n}_{33} } ) }
{\det U^{\n } \det U^{e *} } \bigg ] \nn \\
& = \arg \left [ 
 \dfrac{ R_{12}^{*} ( 1+ R_{23}^{*}) +  |U_{11}^{e} U_{11}^{\n}|^{2}  }
 {  R_{12} (1+ R_{23} ) -  |U_{11}^{e} U_{12}^{\n}|^{2}} \right] 
  + \arg \bigg[ R_{12} (  1 + R_{23} )  \bigg ]  . 
\label{general}
\end{align}

In particular,  where $U_{23}^{e}$ is also neglected, the limit $R_{23} \to 0$ yields 
\begin{align}
\d \simeq \arg \left [ 
 R_{12}  \dfrac{  R_{12}^{*} + |U_{11}^{e} U_{11}^{\n}|^{2}   }{  R_{12} - |U_{11}^{e} U_{12}^{\n}|^{2} } \right]  . 
 \label{dR12}
\end{align}
The neglected matrix elements $U_{13}^{\n}$, $U_{13}^{e}$, and $U_{23}^{e}$ are expected to be small owing to the chiral symmetries. Therefore, this simple expression for the Dirac phase is expected to capture the dominant contribution in many models. 
The mixing matrix for this approximation in the standard parametrization is 
\begin{align}
U = 
\begin{pmatrix}
c_{12}^{e} & -s_{12}^{e} & 0 \\
s_{12}^{e} & c_{12}^{e} & 0 \\
0 & 0 & 1 \\
\end{pmatrix}
\Diag{e^{i \r_{1}}}{e^{ i \r_{2}}}{e^{ i \r_{3}}} 
\begin{pmatrix}
1 & 0 & 0 \\
0 & c_{23}^{\n} & s_{23}^{\n} \\
0 & - s_{23}^{\n} & c_{23}^{\n}
\end{pmatrix}
\begin{pmatrix}
s_{12}^{\n} & s_{12}^{\n} & 0 \\
- s_{12}^{\n} & c_{12}^{\n} & 0 \\
0 & 0 & 1 
\end{pmatrix}  , 
\end{align}
and the CP phase is obtained as
\begin{align}
\d =  \arg \left [  \frac{ e^{-i \rho _1} c^{\nu }_{23} s^e_{12} s^{\nu }_{12} + e^{-i \rho _2} c^e_{12} c^{\nu }_{12} }
{ e^{- i \rho _2} c^{\nu }_{12} c^{\nu }_{23} s^e_{12} - e^{- i \rho _1} c^e_{12} s^{\nu }_{12} } \right ]  \, . 
\end{align}
Since $\r_{3}$ can be easily removed by a rephasing transformation, 
it is clear that the physical CP violation depends only on $\r_{1} - \r_{2}$.
%

\subsection{TBM mixing and CKM-like charged-lepton mixing}

To explore how the above results can be applied in future observations, 
we examine specific numerical examples. Here, let us consider the typical tri-bimaximal (TBM) mixing \cite{Harrison:2002er} together with a CKM-like  mixing of charged leptons,
\begin{align}
U =  U^{e \dg} U_{\rm TBM} \, .
\end{align}
Here, $U^{e}$ and $U_{\rm TBM}$ are taken in their general forms after phase redefinitions 
\begin{align}
U^{e} =
\begin{pmatrix}
\cos \th & e^{- i \phi}  \sin \th & 0 \\
-e^{ i \phi}  \sin \th  &  \cos \th & 0 \\
0 & 0 & 1 \\
\end{pmatrix} , 
~~~
U_{\rm TBM} 
= 
\begin{pmatrix}
 \sqrt{\frac{2}{3}} & \frac{1}{\sqrt{3}} & 0 \\
- \frac{1}{\sqrt{6}} & \frac{1}{\sqrt{3}} & - \frac{1}{\sqrt{2}} \\
 -\frac{1}{\sqrt{6}} & \frac{1}{\sqrt{3}} & \frac{1}{\sqrt{2}} \\
\end{pmatrix} , 
\end{align}
where $\sin \th \simeq 0.22$ and $\phi$ is a CP phase associated with the 1-2 mixing.
The magnitudes of $|U_{\a i}|$ are close to observations of neutrino oscillation: 
\begin{align}
U =  
\begin{pmatrix}
 0.796 +0.090 \, e^{-i \phi } & 0.563 -0.127 \, e^{-i \phi } & 0.156 \, e^{-i \phi } \\
 -0.398 + 0.180 \, e^{i \phi } & 0.563 +0.127 \, e^{i \phi } & -0.690 \\
 -\frac{1}{\sqrt{6}} & \frac{1}{\sqrt{3}} & \frac{1}{\sqrt{2}} \\
\end{pmatrix} . 
\label{numU}
\end{align}

Since we have set $U_{23}^{e} = 0$ in this case, 
the Dirac phase $\delta$ is obtained by calculating $R_{12}$,  namely 
\begin{align}
R_{12} = 
 { U^{e *}_{11}  U^{e *}_{12}  U^{e *}_{33} \over \det U^{e *}  } { U^{\n}_{11} U^{\n}_{12}  U^{\n}_{33} \over \det U^{\n} } 
 =  {1\over 3} e^{ i \phi} \cos \th \sin \th \, .
\end{align}
Therefore, from Eq.~(\ref{dR12}),
\begin{align}
\d = \arg \left [ 
 R_{12}  \dfrac{  R_{12}^{*} + |U_{11}^{e} U_{11}^{\n}|^{2}   }{  R_{12} - |U_{11}^{e} U_{12}^{\n}|^{2} } \right]  
 = \arg  \left[ 
 - e^{i \phi}  \sin  \theta 
\frac{ 2 \cos \theta + e^{-i \phi } \sin \theta }
{   \cos\theta -  e^{i \phi}  \sin  \theta  }\right ] . 
\end{align}
Note that real coefficients are essentially irrelevant for the determination of $\delta$. 
For small $\sin \theta$, one finds $\delta \simeq \phi + \pi$ and the phase can take arbitrary values.
The contribution of a finite $U_{23}^{e}$ is understood from the general expression~(\ref{general}), 
and small $|U_{23}^{e}| \lesssim 0.05$ can be treated as a perturbation.
A more general analysis including finite $U_{13}^{\nu}$ is presented in the subsequent paper \cite{Yang:2026pdm}.

For describing this system, 
the parametrization proposed by Fritzsch and Xing is more suitable  \cite{Fritzsch:1997fw} 
\begin{align}
U' & = 
\begin{pmatrix}
c_{u} & s_{u} & 0 \\
- s_{u} & c_{u} & 0 \\
0 & 0 & 1 \\
\end{pmatrix}
\begin{pmatrix}
e^{- i \d_{\rm FX}} & 0 & 0 \\
0 & c_{q} & s_{q} \\
0 & - s_{q} & c_{q}
\end{pmatrix}
\begin{pmatrix}
c_{d} & - s_{d} & 0 \\
s_{d} & c_{d} & 0 \\
0 & 0 & 1 
\end{pmatrix} \nn \\ 
& = 
\begin{pmatrix}
 s_u s_d c_q + c_u c_d e^{-i \delta _{\text{FX}}} & s_u c_d c_q-c_u s_d e^{-i \delta _{\text{FX}}} & s_u s_q \\
 c_u s_d c_q - s_u c_d e^{-i \delta _{\text{FX}}} & c_u c_d c_q + s_u s_d e^{-i \delta _{\text{FX}}}& c_u s_q \\
 -s_d s_q & -c_d s_q & c_q \\
\end{pmatrix} . 
\label{originalFX}
\end{align}
It decomposes the quark mixings into the lighter 1-2 and the heavier 2-3 mixing, 
based on the characteristic features of the quark hierarchy.  
The CP phase $\delta_{\rm FX}$ is expressed by a rephase invariant, 
and from Eq.~(\ref{numU}), 
it is written without using the $\arg$ function: 
\begin{align}
\d_{\rm FX} =  \arg \left[ {U_{e3} U_{\m 3} U_{\t 1} U_{\t 2} \over U_{\t 3} \det U } \right]
= - \phi \, . 
\end{align}
A rephasing invariant analysis of this FX phase will also be carried out in a future work.

\section{Conclusion}

In this letter, we presented the explicit rephasing transformation
of an arbitrary mixing matrix to the PDG standard form
using arguments of its matrix elements and determinant.
By this procedure, six independent CP phases are identified systematically as combinations of the arguments. 
Similar explicit rephasing transformations will exist in other representations as well, and the transformations between them are carried out simply by the multiplication of diagonal phase matrices.

We also apply this framework to the fermion diagonalization matrices $U^{\nu,e}$ under the approximation $U^{\nu,e}_{13}=0$. 
By partially employing the inversion formula, we factorize all CP phases through the explicit rephasing, thereby revealing the entire CP structure of the mixing matrix.
The observable Dirac phase depends only on two relative phases between the fermion sectors and is determined by fermion-specific invariants. 
An understanding of global phase structures offers a useful formulation for describing the Dirac and Majorana phases, 
enabling a clearer perspective of their correlations and renormalization group equations.  

\section*{Acknowledgment:}
The study is partly supported by the MEXT Leading Initiative for Excellent Young Researchers Grant Number JP2023L0013.


\end{document}